\documentstyle[twocolumn,prl,aps,epsf,floats]{revtex}

\begin{document}
\draft
\def \beq{\begin{equation}}
\def \eeq{\end{equation}}
\def \beqarr{\begin{eqnarray}}
\def \eeqarr{\end{eqnarray}}

\twocolumn[\hsize\textwidth\columnwidth\hsize\csname @twocolumnfalse\endcsname

\title{
Disorder Induced Cluster Formation near First Order Phase Transitions in 
Electronic Systems:
Importance of Long-Range Coulomb Interaction
}

\author{Kun Yang}

\address{
National High Magnetic Field Laboratory and Department of Physics,
Florida State University, Tallahassee, Florida 32306
}

\date{\today}
 
\maketitle

\begin{abstract}

We discuss the effects of fluctuations of the local density of charged
dopants near a first order phase transition in electronic systems, 
that is driven by change of
charge carrier density controlled by doping level.
Using a generalization of the Imry-Ma argument,
we find that the first order transition is rounded by disorder at or below
the lower critical dimension $d_c=3$, when at least one of the two phases has
no screening ability. The increase of
$d_c$ from 2 (as in the random field Ising model)
to 3 is due to the long-range nature
of the Coulomb interaction. This result suggests that large clusters of both 
phases will
appear near such transitions due to disorder, in both two and
three dimensions. Possible implications of our results on manganites and 
underdoped cuprates 
%(where meso-scale cluster formation has been found)
will be discussed. 

\end{abstract}

]

Study of effects of quenched disorder on first order phase transitions has a
long history. The prototype model for such studies is the random field
Ising model (RFIM)\cite{im}. 
At low-temperature the pure Ising model spontaneously
develops magnetization and there are two different phases corresponding to the
two possible directions of the magnetization. A {\em uniform} external field $h$
drives the systems through a first order phase transition at $h_c=0$, where the
magnetization changes discontinuously. If a {\em random} field with zero mean is
applied on the other hand, it would like the local order parameter to point 
in the direction of the local field, thus clusters or domains of both phases
would form, the global magnetization would be destroyed, and the first order 
transition associated with the jump of global magnetization would be rounded.
In order for this to happen however, the energy gain due to forming clusters
with magnetization following the local random field direction, $E_h$,
must be large
enough to overcome the energy penalty associated with the domain walls 
separating the clusters with opposite magnetization, $E_w$. 
Imry and Ma\cite{im}
analyzed the competition between these two effects as a function of the linear
size of the cluster $L$, and found that for systems with dimensionality $d$
less than the lower critical dimension $d_c=2$, $E_h\sim L^{d/2}$ overwhelms 
$E_w\sim L^{d-1}$ for large enough $L$; in this case clusters will form at 
sufficiently large length scales {\em no matter how weak the random field is},
and the first order transition is rounded. For $d > d_c$ on the other hand 
the global magnetization and first order transition remain intact for {\em weak}
random field. For $d=d_c=2$, $E_h$ and $E_w$ scale with $L$ the same way:
$E_h\sim E_w\sim L$. More detailed studies\cite{binder,aw}
indicated that at $d=d_c=2$, 
$E_h$ eventually dominates $E_w$ at large $L$ and destroys the ordered phase
and the associated first order transition. These considerations and conclusions 
have been generalized to generic first order phase transitions\cite{iw}. 

\begin{figure}
\epsfxsize=3.6in
%\centerline{ \epsffile{fig.eps} }
%\vskip -0.7 in
\epsfbox{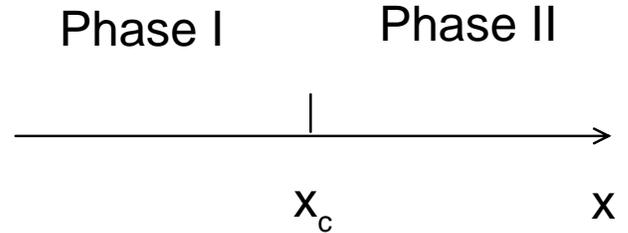}
%\vskip -1.4 in
\caption{
Schematic phase diagram of an electronic system. As the doping level $x$ is
varied, there are two phases separated by a first order phase transition at 
$x=x_c$.
}
\label{fig1}
\end{figure}

Recently there has been considerable interest in the rounding of first order
phase transitions in electronic systems, and the associated cluster formation.
In these systems the first order transition in the absence of disorder
is driven by the change of charge carrier density, which
in turn is controlled by the density of dopants introduced into the system; see
Fig. 1. 
Here the dopants play a dual role. First of all they provide the charge carriers
(conduction electrons or holes), and their density determines the location of
the system in the phase diagram. On the other hand they are also the source 
of disorder, as they distort the local lattice structure and {\em more 
importantly, there is local fluctuation in their density}. 
As we argue below, due to the long-range nature of the Coulomb interaction 
between the dopants and the charge carriers, the effect of dopant density 
fluctuation is much stronger than the random field in the random field Ising
model; the latter only couples to the {\em local} order parameter and thus 
represents short-range interaction. A consequence of that is it increases the 
lower critical dimension 
$d_c$ from 2 to 3; we thus expect in both 2- and 3-dimensional
systems the otherwise first-order
phase transitions are rounded by dopant density fluctuations, and clusters of
both phases will always form near the phase boundary. In the following we
illustrate these points by performing an extension of the original 
Imry-Ma analysis. We will consider two different generalizations
of the Coulomb interactions in 3D to arbitrary dimensions. The first is to 
keep the momentum dependence: $V(q)\sim 1/q^2$, which yields $V^{(1)}(r)\sim r^{2-d}$
in real space (in 2D, $V^{(1)}(r)\sim \log r$); the other is simply using 
$V^{(2)}(r)\sim 1/r$ in all dimensions. We will discuss both cases.

Now assume that the {\em average} dopant density of the system is $\overline{n}
=n_c$, {\em i.e.} at the transition point (or doping level $x=x_c$ in Fig. 1). 
In the {\em absence} of any disorder,
the system will be in one of the two phases depicted in Fig. 1 and there are no
clusters or domains, since generically there is an energy penalty associated 
with the domain walls that is proportional to the area of the walls. We now 
discuss what happens when there is local fluctuation of the dopant density.
Consider a domain of linear size $L$. The average number of dopants in this
domain is $N(L)\sim L^d$, while the typical {\em fluctuation} of $N(L)$ is
$\delta N(L)\sim \sqrt{N(L)}\sim L^{d/2}$. Obviously, the local fluctuation of dopant
density tend to force domains of both phases to form, according to the local
density of dopants; the energy penalty of {\em not} forming domains is dominated
by the electrostatic energy due to the imbalance between the dopant charge and
carrier charge in the regions that are in the ``wrong" phase where the carrier
density cannot reach that of the local dopant density, which is 
\beq
E_c(L)\sim [\delta N(L)]^2 V(L)\sim L^d V(L)
\eeq
per domain. For the two types of generalization of Coulomb interaction, we find
\beq
E_c^{(1)}(L)\sim L^2
\label{c1}
\eeq
and 
\beq
E_c^{(2)}(L)\sim L^{d-1}
\label{c2}
\eeq
respectively. In order for the domain formation to occur, the gain of Coulomb
energy $E_c$ must overcome the domain wall energy penalty 
\beq
E_w\sim L^{d-1}.
\label{wall}
\eeq
Comparing the expression of $E_w$ with those of $E_c^{(1)}$ and $E_c^{(2)}$,
we find for the first generalization of Coulomb interaction, 
$E_c^{(1)}$ dominates
$E_w$ at sufficiently large $L$ for $d < d_c =3$, {\em i.e.}, the lower critical
dimension above which the first order transition remains intact is $d_c=3$; at
3D the effect of the dopant density fluctuation is marginal. For the second
generalization of Coulomb interaction, $E_c^{(2)}$ scales with $L$ exactly the
same way as $E_w$, {\em regardless of the dimensionality d}.
Thus the effects of dopant density fluctuation is always marginal. 

These results
are very different from that of the random field Ising model (RFIM), 
in which case
$d_c=2$; the origin of the fundamental difference lies in the long-range nature
of the Coulomb interaction, which yields very different dependence of $E_c$ on
the domain size $L$, as compared with $E_h\sim L^{d/2}$ of the RFIM.

The discussions presented 
above are quite general. In the following we propose a simple lattice
model that contains the basic ingredients of the physics being considered here,
which can be studied numerically to check the results obtained 
above\cite{note1}:
\beq
H={1\over 2}\sum_{ij}{(\rho_i-n_i)(\rho_j-n_j)\over |{\bf r}_i-{\bf r}_j|}
+{J\over 2}\sum_{<ij>} [1-sgn(\rho_i\cdot\rho_j)].
\eeq
Here $i$ and $j$ are indices of lattice sites of a $d$-dimensional cubic 
lattice, $n_i$ is the quenched background (or dopant) charge which follows some
specific distribution like Gaussian or box distributions, and $\rho_i$ is 
the electron charge which is the dynamical variable of the model, and it is   
varied to minimize $H$. 
The first term represents the $1/r$ Coulomb interaction among the charge
carriers, and the summation is over all pairs of sites in the system; it can
also be replaced by other forms of long-range interaction that is appropriate. 
The second term represents domain wall energy cost and the sum is over nearest 
neighbors; here without losing generality we have assumed the phase boundary 
is located at $\rho_c=0$, thus there is an energy cost of $J$ for each 
nearest neighbor bond
that connects sites with opposite signs of $\rho$, or sites occupied by
different phases (the function $sgn$ returns the sign of its argument).  
It is clear that in the absence of fluctuation in the background (or dopant) 
charge density, {\em i.e.}, $n_i=n$ independent of $i$, we will simply have
$\rho_i=n$ everywhere, there is no competition between the two terms in $H$,
and there is a first order transition at $n=0$ from the phase with $\rho < 0$
to the phase with $\rho > 0$. On the
other hand when $n_i$ follows certain distribution with a finite width, and its
average $\overline{n}$ is close to $0$, these two terms compete with each other.
The first term prefers to have $\rho_i$ match $n_i$ everywhere, thus 
$\rho_i$ with both signs (or both phases) would appear in the system; the
fluctuation of $n_i$ plays a role similar to the random field in the RFIM.
On the 
other hand the second term penalizes near neighbor pairs with opposite signs of
$\rho$; $J$ plays a role similar to the Ising coupling in the RFIM.
Numerical simulation of this and other closely related models are currently
underway\cite{note2}.

A few comments on our results are now in order. 

(i) It is very important in our discussion that the first order transition is
driven by the change of doping level. There are, of course, other first order
transitions in electronic systems that are driven by other mechanisms (like
change of
pressure) {\em without} changing the doping level; our results do not apply to
those cases, as long as the position of the first order transition does not 
involve the doping level.
 
(ii) Within the RFIM, it is known that if there exists power-law correlation 
in the random field, the lower critical dimension $d_c$ may be 
increased\cite{natterman}. The increase of $d_c$ due to long-range Coulomb 
interaction discussed here has some resemblance to this effect, but the precise
mechanism of increase of $d_c$ is different.

(iii) In the case of the RFIM, it has been proven rigorously\cite{aw}
that right at the 
lower critical dimension $d=d_c=2$, the random field destroys the ordered phase,
thus the first order transition is rounded by the cluster or domain 
formation\cite{note4}.
We are unable to provide such a proof in the case we study here. 
However heuristically it is easy to see why at $d=d_c$ disorder effects 
eventually dominate. This is because for a given domain size scale $L$, 
there is 
always some probability that the fluctuation of the number of dopants within
a domain is strong enough to overcome the energy penalty of the 
corresponding domain wall; such probability does {\em not} decrease with
increasing $L$ at $d=d_c$. Therefore domains with arbitrary size will appear
at $d=d_c$.
We thus expect that for the $1/r$ Coulomb
interaction, disorder effects dominate and the first order transitions are
rounded by cluster formation in both $d=2$ and $d=3$.

(iv) Thus far we have not explicitly
discussed the nature of the phases that are involved, and how
they may affect the consideration above. If both of the competing phases are
insulating and have {\em no} screening ability, 
then all of our discussion will just come through with no 
modification\cite{note}. 
The situation is more complicated and interesting when metallic
phases are involved. Metallic phases have screening abilities; they can
maintain charge neutrality of the system at large scales by creating {\em local}
carrier density deviation near a charged impurity to compensate for the its
charge; {\em locally}, the carrier density may deviate from its average value
significantly and may even move across the phase boundary. Thus when {\em both}
phases are metallic, the charge imbalance discussed above may be compensated
for by screening processes, and our conclusions do not apply beyond the 
screening length. 
This is {\em not} the case however when 
{\em one of the two competing phases is insulating}; this is because the 
insulating phase does not have the screening ability, thus cluster formation
cannot be avoided this way when the phase boundary is approached from the
insulating side, 
%and the first order transition is {\em still} rounded.
and the transition from an insulator to a metal is a percolation transition.
Even when both phases have screening abilities, our results are still relevant
when the screening length $\xi$
is long compared to microscopic length scales
(typical for `` poor" metals). In this case
clusters will form at $d=3$ up to size $L\sim \xi$ no matter how weak the 
dopant fluctuation is. 

We emphasize that the considerations presented here are rather generic, and
the results should be relevant to any first order phase transitions driven by
change of doping level in electronic systems. As {\em possible} examples of 
the situation discussed here, we now turn our discussion to two electronic
systems that are of tremendous current interest, namely the manganites with
colossal magneto resistance and high $T_c$ cuprates. 
Both compounds have very rich
phase diagrams, and in particular, there is a transition from an insulating
phase to a metallic phase as dopant concentration is varied (in the latter case
the metallic phase is superconducting at low temperature). 
Recently there has been considerable experimental evidence suggesting that
clusters of both phases co-exist and the system becomes inhomogeneous, 
when the transition is approached\cite{cheong,mydosh,howald,pan,iguchi,lang,renner,greene}.
It has been suggested that such inhomogeneity is induced by disorder near an
otherwise first order transition\cite{moreo,dagotto,burgy,wang,dbm}. 
In the case of the manganites, convincing arguments have been advanced that
the cluster formation and the percolative nature of the metal-insulator
transition may be the origin of the colossal 
magnetoresistance\cite{moreo,dagotto,burgy,dbm}, 
and prediction of possible similar behavior in the cuprates has been made based
on the similarity between these two systems\cite{dbm}.
We note that the manganites are 3D systems, which is {\em above}
the lower critical
dimension of the RFIM, $d_c=2$. The cuprates are quasi-2D systems; however its
large distance behavior must be determined by the 3D nature of the system.
Thus without long-range interactions, the formation of clusters and the 
resultant inhomogeneity due to disorder near a first order transition is 
possible but would {\em not} be guaranteed in either case\cite{note3}. 
The importance of the long-range
nature of the Coulomb interaction, as we discussed here, is to {\em guarantee}
cluster formation and inhomogeneity by pushing the lower critical dimension
up to $d_c=3$. 
%at least up to the scale of
%screening length which is very long in both
%systems. 
Our results are thus consistent with the phenomenology of both the manganites
and cuprates, and point to the generality of cluster formation near first order
phase transitions in electronic systems, which is of central importance in
the theory of 
Dagotto and co-workers on the colossal magnetoresistance in the manganites, 
based on phase separation and cluster formation. 

The author thanks Elbio Dagotto, Sankar Das Sarma, and Vlad Dobrosavljevic for
encouragements, stimulating discussions, and helpful comments on the 
manuscript. He also thanks the Aspen Center for Physics for hospitality, where
this work was initiated during her Workshop on Collective Phenomena in Glassy
Systems. This work was supported by NSF grants No. DMR-9971541 and
DMR-0225698, and the A. P.
Sloan Foundation.


\begin{references}
\bibitem{note1} This model was conceived during stimulating discussions with
Elbio Dagotto.
\bibitem{note} In reality, all experiments are performed at finite temperatures,
and even insulators have some screening ability. However as long as the 
screening length is sufficiently long,
%much longer than other relevant length scales in the system,
our results still apply; see discussions below.
\end{references}
\end{document}